# Technical Report

*Brain Invaders Adaptive versus Non-Adaptive P300 Brain-Computer Interface dataset*

~

E. Vaineau, A. Barachant, A. Andreev, P. Rodrigues, G. Cattan, M. Congedo

GIPSA-lab, CNRS, University Grenoble-Alpes, Grenoble INP.
Address : GIPSA-lab, 11 rue des Mathématiques, Grenoble Campus BP46, F-38402, France

**Abstract** - We describe the experimental procedures for a dataset that we have made publicly available at https://doi.org/10.5281/zenodo.1494163 in *mat* and *csv* formats. This dataset contains electroencephalographic (EEG) recordings of 24 subjects doing a visual P300 Brain-Computer Interface experiment on PC. The visual P300 is an event-related potential elicited by visual stimulation, peaking 240-600 ms after stimulus onset. The experiment was designed in order to compare the use of a P300-based brain-computer interface on a PC with and without adaptive calibration using Riemannian geometry. The brain-computer interface is based on electroencephalography (EEG). EEG data were recorded thanks to 16 electrodes. Data were recorded during an experiment taking place in the GIPSA-lab, Grenoble, France, in 2013 (1). Python code for manipulating the data is available at https://github.com/plcrodrigues/py.BI.EEG.2013-GIPSA. The ID of this dataset is *BI.EEG.2013-GIPSA*.

**Résumé** - Dans ce document, nous décrivons une expérimentation dont les données ont été publiées sur https://doi.org/10.5281/zenodo.1494163 aux formats *mat* et *csv*. Ce jeu de donnée contient les enregistrements électroencéphalographiques (EEG) de 24 sujets durant une expérience sur les interfaces cerveau-ordinateur de type 'P300 visuel'. Le P300 visuel est une perturbation du signal EEG apparaissant 240-600 ms après le début d'une stimulation visuelle. Le but de cette expérience était de comparer l'utilisation d'une interface cerveau-machine (ICM) basée sur le P300, sous PC, avec et sans calibration adaptive en utilisant la géométrie Riemannienne. L'EEG de chaque sujet a été enregistré grâce à 16 électrodes réparties sur la surface du scalp. L'expérience a été menée au GIPSA-lab (Université de Grenoble-Alpes, CNRS, Grenoble-INP) en 2013 (1). Nous fournissons également une implémentation python pour manipuler les données à https://github.com/plcrodrigues/py.BI.EEG.2013-GIPSA. L'identifiant de cette base de données est *BI.EEG.2013-GIPSA*.

## Introduction

The visual P300 is an event-related potential (ERP) elicited by a visual stimulation, peaking 240-600 ms after stimulus onset. The experiment was designed in order to compare the use of a P300-based brain-computer interface on a PC with and without adaptive calibration using Riemannian Geometry (1, 6). The experiment used the *Brain Invaders* P300-based Brain-Computer Interface (BCI) (2,3). For classification purposes the Brain Invaders implemented an online Riemannian Minimum Distance to Mean (RMDM) classifiers (4–7). This experiment features both a training-test (classical) mode of operation and a calibration-less mode of operation (6–8). An example of application of this dataset can be seen in (9).

## Participants

The performance of the BCI with and without adaptive calibration were assessed in both transversal and longitudinal experiments. 24 subjects participated in the transversal experiment (12 females), with mean (sd) age 25.96 (4.46). Among them, seven subjects (four females) with mean (sd) age 27.14 (7.24) participated in the longitudinal experiment. All subjects were volunteers recruited by means of flyers and thanks to the mailing list of the University of Grenoble-Alpes. An interview was conducted with each of them and we excluded from the study the volunteers who were not meeting the criteria presented in the Appendix. The study was approved by the Ethical Committee of the University of Grenoble Alpes (Comité d'Ethique pour la Recherche Non-Interventionnelle). All participants provided written informed consent confirming the notification of the experimental process, the data management procedures and the right to withdraw from the experiment at any moment.

## Material

Stimulus was displayed on a ViewSonic (California, US) screen with length 22''. EEG signals were acquired by means of a research-grade amplifier (g.USBamp, g.tec, Schiedlberg, Austria) and the g.GAMMAcap (g.tec, Schiedlberg, Austria) equipped with 16 wet Silver/Silver Chloride electrodes, placed according to the 10-20 international system (**Figure 1**). The locations of the electrodes were FP1, FP2, F5, AFZ, F6, T7, CZ, T8, P7, P3, PZ, P4, P8, O1, OZ and O2 (**Figure 1**). The reference was placed on the left earlobe and the ground at the FZ scalp location. The amplifier was linked by USB connection to the PC where the data were acquired by means of the software OpenVibe (10,11). Data were acquired with no digital filter applied and a sampling frequency of 512 samples per second. For ensuing analysis, the application tagged the EEG using USB. The tags were sent by the application to the amplifier through the USB port of the PC. It was then recorded along with the EEG signal as a supplementary channel. The tagging process was the same in all experimental conditions, that is, the jitter and latency of the tagging was identical with and without adaptive calibration. This allows comparing the resulting ERP between the experimental conditions, without correcting the tagging latency (12).

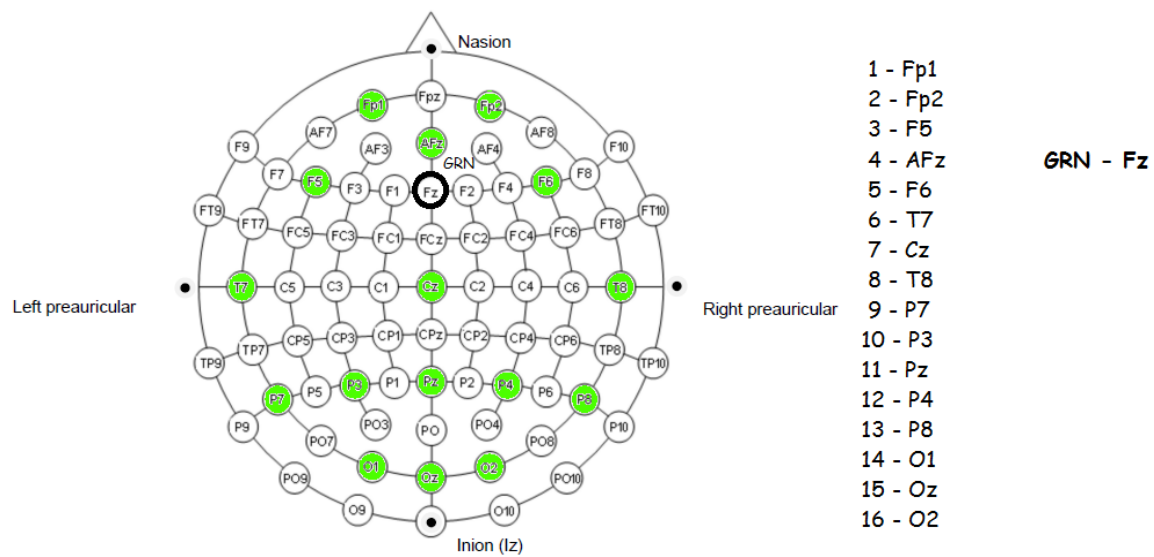

**Figure 1.** In green, the 16 electrodes placed according to the 10-20 international system. We used FZ (circled in black) as ground and the left earlobe as a reference.

**Procedures**

For all subjects, the experiment took place in a small room with a surface of four meters square, containing the ViewSonic (California, US) screen and all the required hardware materials for acquiring the EEG data. The subject was sitting at a distance of 75 to 115 cm from the screen. The EEG headset was placed on all subjects, and the integrity of the recording pipeline was checked by performing preliminary tests, consisting in recording visible signals such as eye blinks. The experimenter controlled the session from an adjacent room equipped with a one-way glass window.

Subjects 1 to 7 participated in eight sessions, run on different days. Subjects 8 to 24 participated in one session. Each session consisted in *two runs*, one in a *Non-Adaptive* (classical) and one in an *Adaptive* (calibration less) mode of operation. The order of the runs was randomised for each session. This design allows the use of exact randomisation testing for testing hypotheses (13). In both runs there was a *Training (calibration) phase* and an *Online phase*, always passed in this order. In the non-Adaptive run the data from the Training phase was used for calibrating the classifier used during the Online phase using the training-test version of the MDM algorithm (5,6). In the Adaptive session, the data from the training phase was not used at all, instead the classifier was initialised with generic class geometric means (the class grand average estimated on a database of subjects participating to a previous Brain Invaders experiment) and continuously adapted to the incoming data using the Riemannian method explained in (6). Subjects were completely blind to the mode of operation and the two runs appeared to them identical.

The interface of Brain Invaders is compounded by 36 symbols distributed in 12 groups. In the Brain Invaders P300 paradigm, a *repetition* is composed of 12 flashes (*i.e.*, one for each group), of which two include the Target symbol (*Target* flashes) and 10 do not (*non-Target* flash) - **Figure 2**. Please see (2,14) for a full description of this paradigm. For this experiment, in the Training phases the number of flashes is fixed to 80 Target flashes and 400 non-Target flashes. In the Online phases, the number of Target and non-Target still are in a ratio one-to-five. However their number is variable since the number of repetitions needed to destroy the target in the Brain Invaders BCI video game depends on the user's performance (6,8). In any case, since the classes are unbalanced, an appropriate score must be used for quantifying the performance of classification methods, such as the Balanced Accuracy (BA):

$$BA = \frac{1}{2}(\frac{A}{A+B} + \frac{C}{C+D}),$$

where A and B (resp. C and D) stands for the number of correctly and non-correctly classified flashes of non-Target (resp. Target) group.

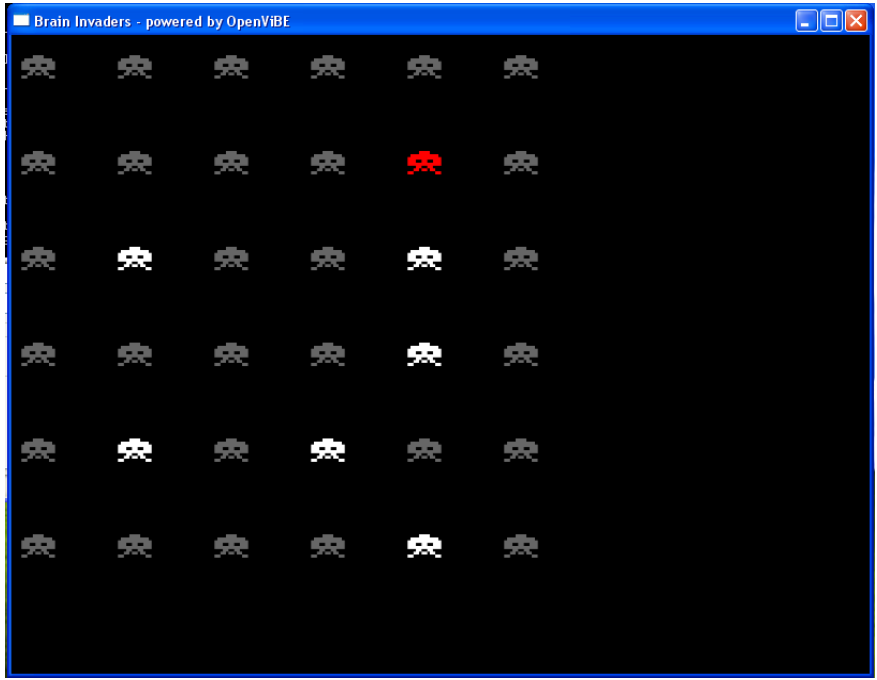

**Figure 2.** Interface of Brain Invaders at the moment where a group of six non-Target symbols flash (in white). The red symbol is the Target. The non-Targets which are not flashing are in grey.

**Organization of the Dataset**

For each subject we provide a *zip* file containing the complete recording of the experiment. This archive contains a folder for each session passed by the subject. In each session folder, there are four *mat* (MathWorks, Natick, US) and one *meta.yml* files. A *mat* file is produced for each run in a session. The *meta.yml* files reports the experimental condition (adaptive, non-adaptive) and the phase (training, online) associated to a *mat* file. Note that all the *mat* files are also provided in *csv* format. Note also that the previous version of the *database* (15) already provides the *gdf* version of the *mat* files.

We supply an online and open-source example working with Python (9) and using the analysis framework MNE (16,17) and MOABB (18,19), a comprehensive benchmark framework for testing popular BCI classification algorithms. This example shows how to download the data and classify 1s non-Target and Target epochs of signals using the xDAWN algorithm (20). This database has been used in the development of the common spatiotemporal pattern method for estimating ERPs (21), which is an extension of the xDAWN algorithm.

# Appendix: Criteria of Participation

**Inclusion criteria**

| Age between 20 and 30 years old | yes ☐ no ☐ Age:… |
|---|---|

**Exclusion criteria**

| History of neural pathology such as epilepsies or headache? | yes ☐ no ☐ |
|---|---|
| History of trauma with loss of consciousness? | yes ☐ no ☐ |
| Diabetes, cardiac pathology or immunodeficiency. | yes ☐ no ☐ |
| Medical treatment susceptible to modulate brain or cardiac activity. | yes ☐ no ☐ |
| Recently consumed drugs, alcohol or doping products. | yes ☐ no ☐ |
| Sensorial or motor handicap such as colour blindness. | yes ☐ no ☐ |

# References

1. Congedo M. EEG Source Analysis [Internet] [Habilitation à diriger des recherches]. Université de Grenoble; 2013. Available from: https://tel.archives-ouvertes.fr/tel-00880483

2. Congedo M, Goyat M, Tarrin N, Ionescu G, Varnet L, Rivet B, et al. "Brain Invaders": a prototype of an open-source P300- based video game working with the OpenViBE platform. In: 5th International Brain-Computer Interface Conference 2011 (BCI 2011) [Internet]. 2011. p. 280–3. Available from: https://hal.archives-ouvertes.fr/hal-00641412/document

3. Korczowski L, Barachant A, Andreev A, Jutten C, Congedo M. "Brain Invaders 2" : an open source Plug & Play multi-user BCI videogame. In: 6th International Brain-Computer Interface Meeting (BCI Meeting 2016) [Internet]. Pacific Grove, CA, United States: BCI Society; 2016. p. 10.3217/978-3-85125-467-9. Available from: https://hal.archives-ouvertes.fr/hal-01318726

4. Barachant A, Bonnet S, Congedo M, Jutten C. Classification of covariance matrices using a Riemannian-based kernel for BCI applications. Neurocomputing. 2013 Jul 18;112:172–8.

5. Barachant A, Bonnet S, Congedo M, Jutten C. Multiclass brain-computer interface classification by Riemannian geometry. IEEE Trans Biomed Eng. 2012 Apr;59(4):920–8.

6. Barachant A, Congedo M. A Plug&Play P300 BCI Using Information Geometry. ArXiv14090107 Cs Stat [Internet]. 2014 Aug 30 [cited 2017 Apr 27]; Available from: http://arxiv.org/abs/1409.0107

7. Congedo M, Barachant A, Bhatia R. Riemannian geometry for EEG-based brain-computer interfaces; a primer and a review. Brain-Comput Interfaces. 2017;4(3):155–74.

8. Congedo M, Barachant A, Andreev A. A New Generation of Brain-Computer Interface Based on Riemannian Geometry. ArXiv13108115 Cs Math [Internet]. 2013 Oct 30 [cited 2019 Apr 10]; Available from: http://arxiv.org/abs/1310.8115

9. Rodrigues P. Codes for working with the "Brain Invaders 2013" dataset developed at the GIPSA-lab [Internet]. 2019. Available from: https://github.com/plcrodrigues/py.BI.EEG.2013-GIPSA

10. Renard Y, Lotte F, Gibert G, Congedo M, Maby E, Delannoy V, et al. OpenViBE: An Open-Source Software Platform to Design, Test, and Use Brain–Computer Interfaces in Real and Virtual Environments. Presence Teleoperators Virtual Environ. 2010 Feb 1;19(1):35–53.

11. Arrouët C, Congedo M, Marvie J-E, Lamarche F, Lécuyer A, Arnaldi B. Open-ViBE: A Three Dimensional Platform for Real-Time Neuroscience. J Neurother. 2005 Jul 8;9(1):3–25.

12. Cattan G, Andreev A, Maureille B, Congedo M. Analysis of tagging latency when comparing event-related potentials [Internet]. Gipsa-Lab ; IHMTEK; 2018 Dec. Available from: https://hal.archives-ouvertes.fr/hal-01947551

13. Edgington ES. Randomization tests. New York: Marcel Dekker; 1980.

14. Andreev A, Barachant A, Lotte F, Congedo M. Recreational Applications of OpenViBE: Brain Invaders and Use-the-Force [Internet]. Vol. chap. 14. John Wiley ; Sons; 2016. Available from: https://hal.archives-ouvertes.fr/hal-01366873/document

15. Erwan Vaineau, Alexandre Barachant, Anton Andreev, Marco Congedo. Brain Invaders 2013a [Internet]. Zenodo; 2018 [cited 2019 Apr 17]. Available from: https://doi.org/10.5281/zenodo.1494240

16. Gramfort A, Luessi M, Larson E, Engemann DA, Strohmeier D, Brodbeck C, et al. MNE software for processing MEG and EEG data. NeuroImage. 2014 Feb 1;86:446–60.

17. Gramfort A, Luessi M, Larson E, Engemann DA, Strohmeier D, Brodbeck C, et al. MEG and EEG data analysis with MNE-Python. Front Neurosci [Internet]. 2013;7. Available from: https://www.frontiersin.org/articles/10.3389/fnins.2013.00267/full

18. Barachant A. Mother of All BCI Benchmarks. [Internet]. NeuroTechX; 2017. Available from: https://github.com/NeuroTechX/moabb

19. Jayaram V, Barachant A. MOABB: trustworthy algorithm benchmarking for BCIs. J Neural Eng. 2018 Sep 25;15(6):066011.

20. Rivet B, Souloumiac A. Optimal linear spatial filters for event-related potentials based on a spatio-temporal model: Asymptotical performance analysis. Signal Process. 2013 Feb;93(2):387–98.

21. Congedo M, Korczowski L, Delorme A, Lopes Da Silva F. Spatio-temporal common pattern: A companion method for ERP analysis in the time domain. J Neurosci Methods. 2016;267:74–88.